\documentclass{aa}
\usepackage[usenames,dvipsnames]{xcolor}
\usepackage{natbib}
\usepackage{xspace}
\usepackage{amssymb}
\usepackage{amsmath}
\usepackage{graphicx}
\usepackage{hyperref}

\newcommand{\sigwind}{\dot{\Sigma}_{\mathrm{wind}}} 
\newcommand{\leverarm}{\frac{r_{\mathrm{A}}}{r_{\mathrm{F}}}} 
\begin{document}
\title{Effect of wind-driven accretion on planetary migration}  \titlerunning{Winds and migration}
\authorrunning{Kimmig, Dullemond \& Kley} 
\author{C.N.~Kimmig$^{1}$, C.P.~Dullemond$^{1}$, W.~Kley$^{2}$}
\institute{
  (1) Zentrum f\"ur Astronomie, Heidelberg University, Albert-Ueberle-Str.~2, 69120 Heidelberg, Germany\\
  (2) Institut f\"ur Astronomie und Astrophysik, Universit\"at T\"ubingen, Auf der Morgenstelle 10, 72076 T\"ubingen, Germany
} \date{\today}

\abstract{ {\em Context:}
  Planetary migration is a key link between planet formation models and observed
  exoplanet statistics. So far the theory of planetary migration has focused on
  the interaction of one or more planets with an inviscid or viscously evolving
  gaseous disk. Turbulent viscosity is thought to be the main driver of secular
  evolution of the disk, and it is known to affect the migration process for
  intermediate to high-mass planets. Recently, however, the topic of wind-driven
  accretion is experiencing a renaissance, now that as evidence is mounting that
  protoplanetary disks may be less turbulent than previously thought, and 3-D
  non-ideal magnetohydrodynamic modeling of the wind-launching process is
  maturing.\\
  {\em Aim:} We wish to investigate how wind-driven accretion may affect
  planetary migration. We aim for a qualitative exploration of the main effects,
  rather than a quantitative prediction.\\
  {\em Methods:} We perform 2-D hydrodynamic planet-disk interaction simulations
  with the FARGO3D code in the $(r,\phi)$-plane. The vertical coordinate in the
  disk, and the launching of the wind, are not treated explicitly. Instead, the
  torque of the wind onto the disk is treated using a simple 2-parameter formula.
  The parameters are: the wind mass loss rate and the lever arm.\\
  {\em Results:} We find that the wind-driven accretion process has a different
  way of replenishing the co-orbital region than the viscous accretion process.
  The former always injects mass from the outer edge of the co-orbital region,
  and always removes mass from the inner edge, while the latter injects or
  removes mass from the co-orbital region depending on the radial density
  gradients in the disk. As a consequence the migration behavior can differ very
  much, and under certain conditions it can drive rapid type-III-like outward
  migration. We derive an analytic expression for the parameters under
  which this outward migration occurs.\\
  {\em Conclusion:} If wind-driven accretion plays a role in the secular
  evolution of protoplanetary disks, planetary migration studies have to include
  this process as well, because it can strongly affect the resulting migration
  rate and migration direction.}

\maketitle


\section{Introduction}
Planetary migration is an integral part of the theory of planet formation. A
planet, once it is formed, can migrate to vastly different radial locations, due
to its gravitational interaction with the protoplanetary disk. As a result, the
orbital elements of observed exoplanets may not reflect the location where they
were born. Any model prediction of exoplanetary orbital statistics must
therefore include a treatment of the migration process.

Unfortunately, the process of planetary migration is a complex affair
\citep[see, e.g.,][and references therein]{2012ARA&A..50..211K}. While the effect of
the Lindblad torques is fairly well understood \citep{2002ApJ...565.1257T}, the
torques exerted by the gas in the co-orbital region has turned out to be hard to
predict. They depend on the entropy gradient \citep{2008ApJ...672.1054B}, the
radiative cooling efficiency \citep{2008A&A...487L...9K}, the planet mass
\citep{1997Icar..126..261W}, the viscosity of the disk
\citep{2001ApJ...558..453M}, and even on the motion of the planet itself
\citep{2003ApJ...588..494M}.

Among the above mentioned effects, the role of turbulent viscosity needs
particular scrutiny, because in recent years doubts have been raised about whether
protoplanetary disks are indeed as turbulent as they were believed to be.
Evidence against strong turbulence comes from the velocity dispersion inferred
from CO lines \citep{2015ApJ...813...99F, 2018ApJ...856..117F}, the observed
small scale height of the dust rings in HL Tau \citep{2016ApJ...816...25P},
as well as from the comparison of planet-disk interaction models with recent ALMA
observations \citep{2018ApJ...869L..47Z}.
Also theoretical considerations about the degree of
ionization of the gas in such disks have suggested that the magnetorotational
instability, which is a potent driver of turbulence, may not be able to operate
in the disk, except in the very upper layers \citep{1996ApJ...457..355G}.
Instead, a magnetocentrifugal wind is likely to be launched \citep{2013ApJ...769...76B},
which in turn exerts a torque back onto the disk, and thus drives inward gas motion
in the upper layers of the disk, or in other words: causes accretion within the disk
\citep{2013ApJ...775...73S}.

If wind-driven accretion becomes the new paradigm and replaces the classic
viscous disk theory, then also the theory of planetary migration has to be
reconsidered. In the viscous disk picture, the turbulent viscosity has a
strong influence on the planetary migration \citep{2001ApJ...558..453M}.
When a planet opens a gap, the
viscous evolution of the disk acts against this by feeding gas from both the
inner and the outer disk back into the gap. The depth of the gap, and therefore
the amount of material in the co-orbital (corotation) region, thus depends on
the equilibrium between these two counteracting effects. Weak turbulence leads
to a deep gap and a weak corotation torque. Strong turbulence, on the other
hand, keeps feeding gas into the corotation region, leading to a substantial
corotation torque.

In contrast to the viscous disk theory, in the wind-driven accretion picture the
radial inward motion of the gas in the disk is much more laminar. Accretion will
be only inward, irrespective of the radial gradients of the density. This is
because according to the classic \citet{1982MNRAS.199..883B} model, the
acceleration of the wind is driven by the injection of angular momentum into the
wind; angular momentum that is extracted from the disk.

If we now insert a gap-opening planet into such a disk, the wind-driven gas
motions in the disk will inject mass into the gap only from the outside.  At the
inner edge of the gap, in stark contrast to the viscous disk model, the
wind-driven accretion will remove gas from the gap. In mathematical terms the
difference is that the viscous model drives accretion as a kind of diffusion
process (which can transport gas both inward and outward, dependent on the
density gradient), while the wind-driven model drives accretion as an
advection process (which transports gas only inward).

In this paper we wish to explore how the wind-driven accretion process affects
the migration of a planet. The purpose is to gain understanding, not to obtain
quantitative numbers. We therefore deliberately keep the description of the
wind-driven accretion process extremely simple, and focus on the gas motions
in the plane of the disk.

The structure of this paper is as follows: we first describe the model
assumptions, the equations we solve, and the numerical methods we employ
(Section~\ref{sec-model}). Then we describe the
setup of the simulations in Section~\ref{sec-setup} and show the results of a series of model
calculations we performed in Section~\ref{sec-results}. We discuss the meaning of
these results, and the caveats of the model in Section~\ref{sec-discussion},
and finish with a conclusion (Section~\ref{sec-conclusion}).

\section{Model}\label{sec-model}
\begin{figure}
  \centerline{\includegraphics[width=0.5\textwidth]{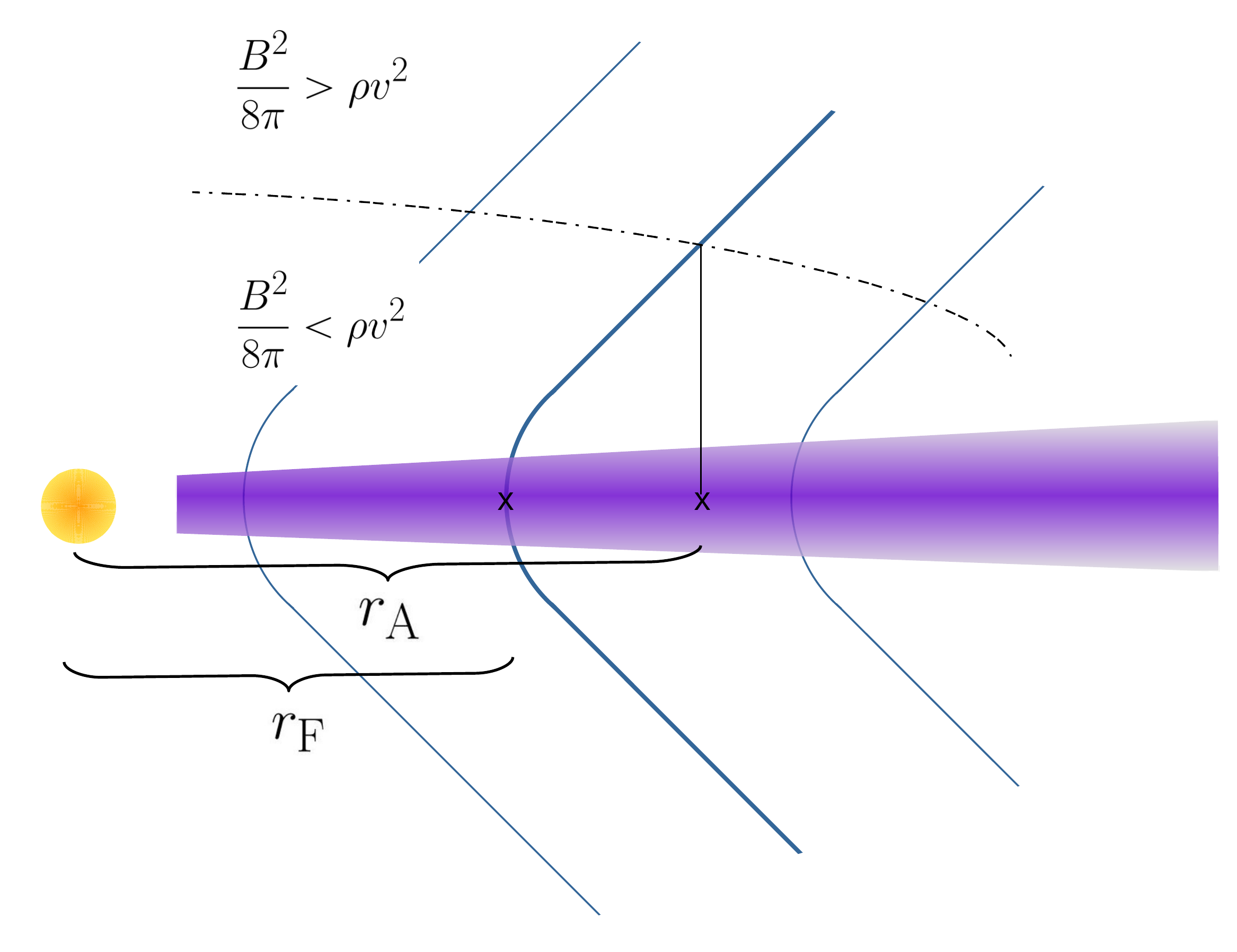}}
  \caption{\label{fig-alfven-surface} Configuration of the magnetic
  field in the \citet{1982MNRAS.199..883B} model (schematic).
  The blue lines represent the magnetic field lines, the dashed line marks the
  Alfv\'{e}n surface and $r_\mathrm{A}$ and $r_\mathrm{F}$ represent
  the Alfv\'{e}n point and the foot point of the (thick) magnetic field line,
  respectively. Inspired by \citet{1996astro.ph..2022S}.
  }
\end{figure}

For the hydrodynamic simulations we use the code FARGO3D by
\citet{2016ApJS.223...11B}. We investigate the effect of the magneto-centrifugal
wind in a two dimensional disk: radial and azimuthal. For simplification, we
implement only the effect of the magnetic wind rather than the magnetic
field itself. The wind is magnetocentrifugally accelerated
\citep{1982MNRAS.199..883B}, meaning that it extracts net angular momentum from
the remaining material in the disk, and thus effects a radial inward drift of
the gas \citep[e.g.,][]{2008NewAR..52...42F, 2016ApJ...821...80B,
  2016A&A...596A..74S}. The resulting radial velocity $v_r$ can then be written
as
\begin{equation} \label{eq-radvelo}
v_r = -2r \ \frac{\sigwind}{\Sigma} \ (\lambda -1),
\end{equation}
with the convention that a negative velocity implies an inward flow. Here,
$\sigwind$ is the mass outflow from the disk, $\Sigma$ the surface density and
\begin{equation}
\lambda = \left(\leverarm\right)^2,
\end{equation}
the lever arm of a magnetic field with $r_\mathrm{A}$ and $r_\mathrm{F}$ being
the Alfv\'{e}n point and the foot point of the magnetic field line,
respectively (see Fig.~\ref{fig-alfven-surface}).

Two parameters determine the strength of the magnetic wind: the magnetic lever
arm $\lambda$ and the mass outflow $\sigwind$. As an assumption for the
parameters, we keep the lever arm constant throughout the disk while we scale
the mass outflow rate $\sigwind$ proportional to the surface density
\begin{equation} \label{eq-massloss}
\sigwind = b \ \frac{\Omega_{\mathrm{K}}}{2 \pi} \Sigma,
\end{equation}
with $\Omega_{\mathrm{K}}$ as Keplerian angular velocity. The mass loss
parameter $b$ is
defined such that it denotes the proportional outflow from the disk per orbital
period. With this parametrization, Eq.~(\ref{eq-radvelo}) turns into
\begin{equation}\label{eq-radvelo-with-b}
v_r = - \frac{\Omega_{\mathrm{K}} r}{\pi} \ b \ (\lambda -1).
\end{equation}
Hence, in our model we have $v_r\propto r^{-0.5}$. This is steeper than
  the velocity profiles in the models by \citet{2017ApJ...845...31H} and
  \citet{2016A&A...596A..74S}, for which they approximately finds $v_r\sim
  r^{-0.215}$ and $v_r\sim r^{-0.25}$, respectively.

We implement this inward drift as an azimuthal torque density on
the disk to decelerate the gas:
\begin{equation}
\begin{split}
  \Gamma &= \sigwind\ \Omega_{\mathrm{K}} r^2 \ (\lambda -1)\\
  &=  -\frac{\Omega_{\mathrm{K}}^2r^2}{2\pi}\ b \Sigma \ (\lambda-1).
\end{split}
\end{equation}
In addition to this torque, we add $-\sigwind$ as a sink term in the
continuity equation.

For a steady radial flow, we need the accretion rate of the gas within the disk
to be constant. The accretion rate $\dot{M} = - 2 \pi r^2 v_r \Sigma$ depends on
the radius $\dot{M} \propto r^{1/2 - p}$ with the assumption of a power law for
the surface density $\Sigma(r) = \Sigma_0 \cdot r^{-p}$. Therefore, a
steady-state solution is possible for $p = 1/2$.

\section{Setup}\label{sec-setup}
To illustrate the effect of the wind-driven torque on planetary migration, we
place a $M_{\mathrm{disk}}=10^{-2}\,M_{\odot}$ disk around a solar mass
star. The disk has a surface density profile $\Sigma\propto r^{-1/2}$ and spans
the radial range between $0.52\,\mathrm{au}$ and $26\,\mathrm{au}$.
With this disk mass, powerlaw and inner/outer boundary, we have
$\Sigma(5.2\mathrm{au})=70.4\,\mathrm{g/cm^2}$.
The radial
temperature structure of the disk is taken to be such, that the aspect ratio 
of the disk is $H_{\mathrm{p}}/r=0.05$, where
$H_{\mathrm{p}}=c_{\mathrm{s}}/\Omega_{\mathrm{K}}$, with $c_{\mathrm{s}}$ the
isothermal sound speed. We set the turbulent viscosity parameter
to $\alpha_{\mathrm{turb}}=0$ \citep{1973A&A....24..337S}. We insert a
planet at $r=r_0=5.2\,\mathrm{au}$,
and during the first ten orbits we allow the mass of the planet to grow linearly
from 0 to the final planet mass $M_\mathrm{p}$. We choose two planetary masses:
$M_\mathrm{p}=100\,M_{\oplus}$ (about a Saturn mass), and
$M_\mathrm{p}=1M_{\mathrm{Jup}}$. These amount to
$q=M_\mathrm{p}/M_{*}$ ratios of $3\times 10^{-4}$ and $10^{-3}$. The
smoothing length of the planet potential is $s=0.6~H_\mathrm{p}$. The
simulations with FARGO3D are dimensionless, so that the results can be scaled to
other stellar masses, provided $q$ stays the same. 

For the $(r,\phi)$ grid we choose 512$\times$656 grid cells. The radial grid is
logarithmically spaced. We run the model for 4000 orbits at $r=r_0$. We employ
the GPU accelerated mode of FARGO3D. For the boundary conditions we choose the
{\tt KEPLERIAN2DDENS} option of FARGO3D, which is an open boundary condition
where the surface density is extrapolated with a powerlaw slope to the ghost
cells. As initial condition for the radial velocity $v_r$ we use
Eq.~(\ref{eq-radvelo-with-b}). The initial condition for the azimuthal velocity
$v_\phi$ is set to Kepler rotation corrected by the pressure gradient.

We vary the mass loss parameter as
$b=10^{-6},\;10^{-5},\;10^{-4},\;10^{-3},\;10^{-2}$. The fiducial value of the
magnetic lever arm is taken to be $\lambda=2.25$
(i.e.,~$r_{\mathrm{A}}/r_{\mathrm{F}}=1.5$).
To focus on an important parameter regime, we add simulations
with $b=2.15\times 10^{-4},\;3.16\times 10^{-4},\;4.64\times 10^{-4},\;6.81\times 10^{-4},\;1.47\times 10^{-3},\;3.16\times 10^{-3}$
for the Saturn planet and $b=3.16\times 10^{-4},\;6.81\times 10^{-4},\;1.47\times 10^{-3},\;2.15\times 10^{-3},\;3.16\times 10^{-3}$
for the Jupiter planet.
We also experiment with
$\lambda=9,\;36,\;81$ (i.e.,~$r_{\mathrm{A}}/r_{\mathrm{F}}=3,\;6,\;9$,
respectively).

For completeness, we can compute the accretion rate $\dot M$ for
our models, given our surface density profile $\Sigma(r)$ and the equation
for the radial velocity (Eq.~\ref{eq-radvelo-with-b}):
\begin{equation}
\begin{split}
\dot M &\equiv 2\pi\Sigma r (-v_r) = 2\Sigma_0r_0^{0.5}\sqrt{GM_{*}}\,b(\lambda-1)\\
&=2.27\times 10^{-4}\,b(\lambda-1)\;M_{\odot}/\mathrm{yr}
\end{split}
\end{equation}
For $\lambda=2.25$ and $b=10^{-4}$ this yields $\dot M=2.8\times
10^{-8}\;M_{\odot}/\mathrm{yr}$.
Comparing this to the viscous disk model with $\dot M = 3 \pi \nu \Sigma$, we would
arrive at a $\alpha$-value of $\alpha = 10^{-2}$.

To test our modifications to FARGO3D, we ran a model without planet, but with a
wind. For the simplified case of torque without corresponding mass
loss, we can test the numerical result against the analytic solution for $v_r$
(Eq.~\ref{eq-radvelo-with-b}). We find a good match.
In the other limiting case of mass loss without corresponding torque
(i.e., $\lambda=1$) an analytic solution of an exponentially dropping
surface density can be found. Also here the numerical result matches it well.

\section{Results}\label{sec-results}
\subsection{Reference case: no wind}
In order to be able to compare our results to a reference case later on, we
first run simulations for the two different planet masses in a non-viscous disk
without magnetic winds. Without viscosity and magnetic winds, the planets should
not migrate.

\begin{figure}
  \centerline{\includegraphics[width=0.5\textwidth]{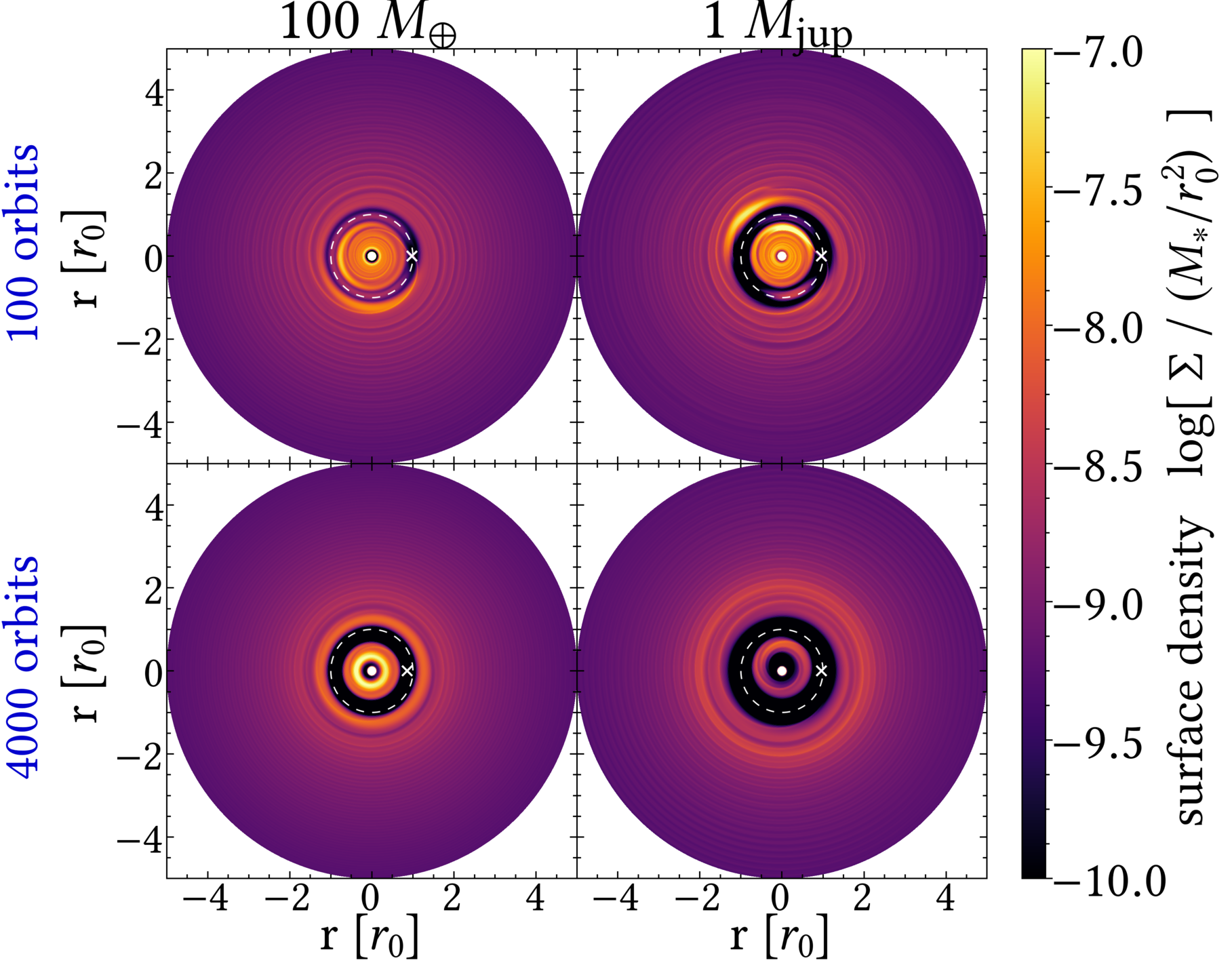}}
  \caption{\label{fig-refcase-surfdens}Results of the inviscid reference models
    without wind-driven accretion: the surface density of the disk for the two
    different planet masses, at 100 and 4000 orbits. The white cross
    denotes the current position of the planet, the dashed white circle marks
    its initial semi-major axis at $r_0 = 5.2~\mathrm{au}$.
  }
\end{figure}

\begin{figure}
  \centerline{\includegraphics[width=0.5\textwidth]{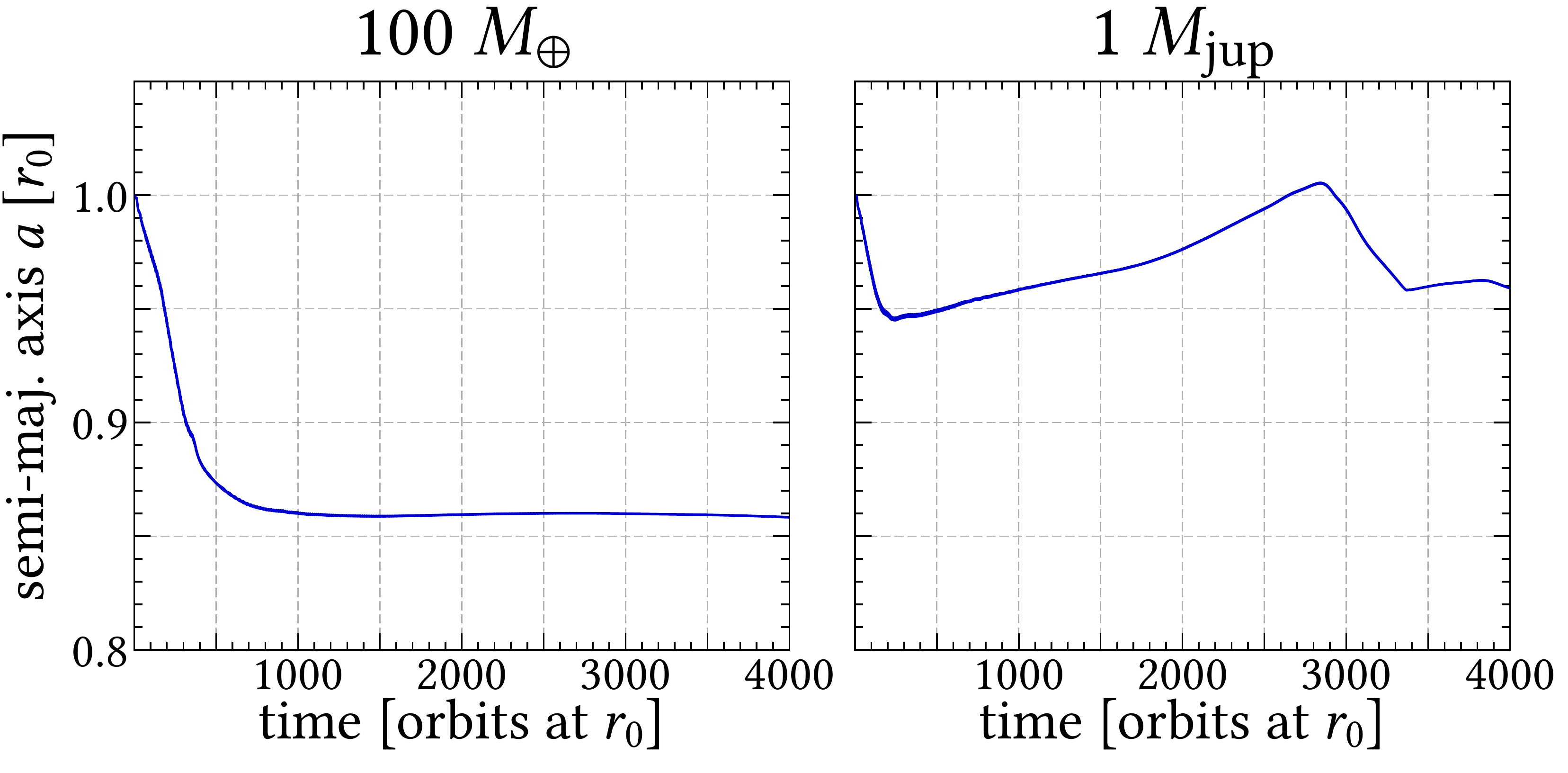}}
  \caption{\label{fig-refcase-migration}Results of the inviscid reference models
    without wind-driven accretion: shown is the time evolution of the
    semi-major axes of the embedded planets.}
\end{figure}

Due to the low viscosity, the planets open a gap in the surface density
(see Fig.~\ref{fig-refcase-surfdens}). This gap formation takes longer
for smaller planet masses. Although the planets slightly change their
semi-major axis (see Fig.~\ref{fig-refcase-migration}), they do not migrate
significantly without viscosity and magnetic winds.

The Saturn-like planet ($M_{\mathrm{p}}=100\,M_\oplus$) migrates in the first 500
orbits, then stalls. This planet takes longer to clear a gap than the
Jupiter-like planet (Fig.~\ref{fig-refcase-surfdens}).
The gas exerts torques on
the planet and it migrates first inward. Once the planet has opened a
gap, the corotation and Lindblad torques are
reduced. As a result, the planet slows down and comes
to a halt. This halting of migration in low-viscosity disks is a known
phenomenon: see, e.g., \citet{2002ApJ...572..566R}, \citet{2009ApJ...690L..52L}
and \citet{ApJ2010...712..198Y}.

For the Jupiter mass planet ($M_{\mathrm{p}}=300\,M_\oplus$), the trend looks
more complicated. In the first one hundred orbits the planet migrates inward,
for the same reason as the Saturn-like planet. It then migrates outward for a
period of 2500 orbits, migrates back inward for 500 orbits, and stalls
there. This migration takes place only in a small scope of
$0.05\,r_0=0.26\,\mathrm{au}$. A possible explanation for this unexpected
behavior is interaction with vortices that form in the disk, which can
be seen in Fig.~\ref{fig-refcase-surfdens}. Such vortices were also
found by \citet{2003ApJ...596L..91K}. The vortices result from
perturbations from the planet in the surface density of the disk, potentially
caused by the short ramp-up time of the planet mass within ten orbits.
These vortices would diffuse in a viscous disk on short time
scales. In the non-viscous disk, however, they stay much longer and accelerate
or decelerate the planet. They therefore influence its semi-major axis. Another
factor influencing its semi-major axis could be the inner disk. In the surface
density (Fig.~\ref{fig-refcase-surfdens}), the inner disk becomes eccentric
due to the open inner boundary condition.
The planet's orbit, however, stays nearly circular: it's eccentricity
staying below about 0.02.
The eccentricity of the inner disk causes that on one
side, the planet is closer to the inner disk than on the other side. This exerts
an asymmetric torque on the planet which either accelerates or decelerates the
planet, causing it to migrate. Thus, the inner disk could also explain the
unexpected behavior of the planet.

\subsection{Model with wind-driven accretion}
\begin{figure*}
  \centerline{\includegraphics[width=0.8\textwidth]{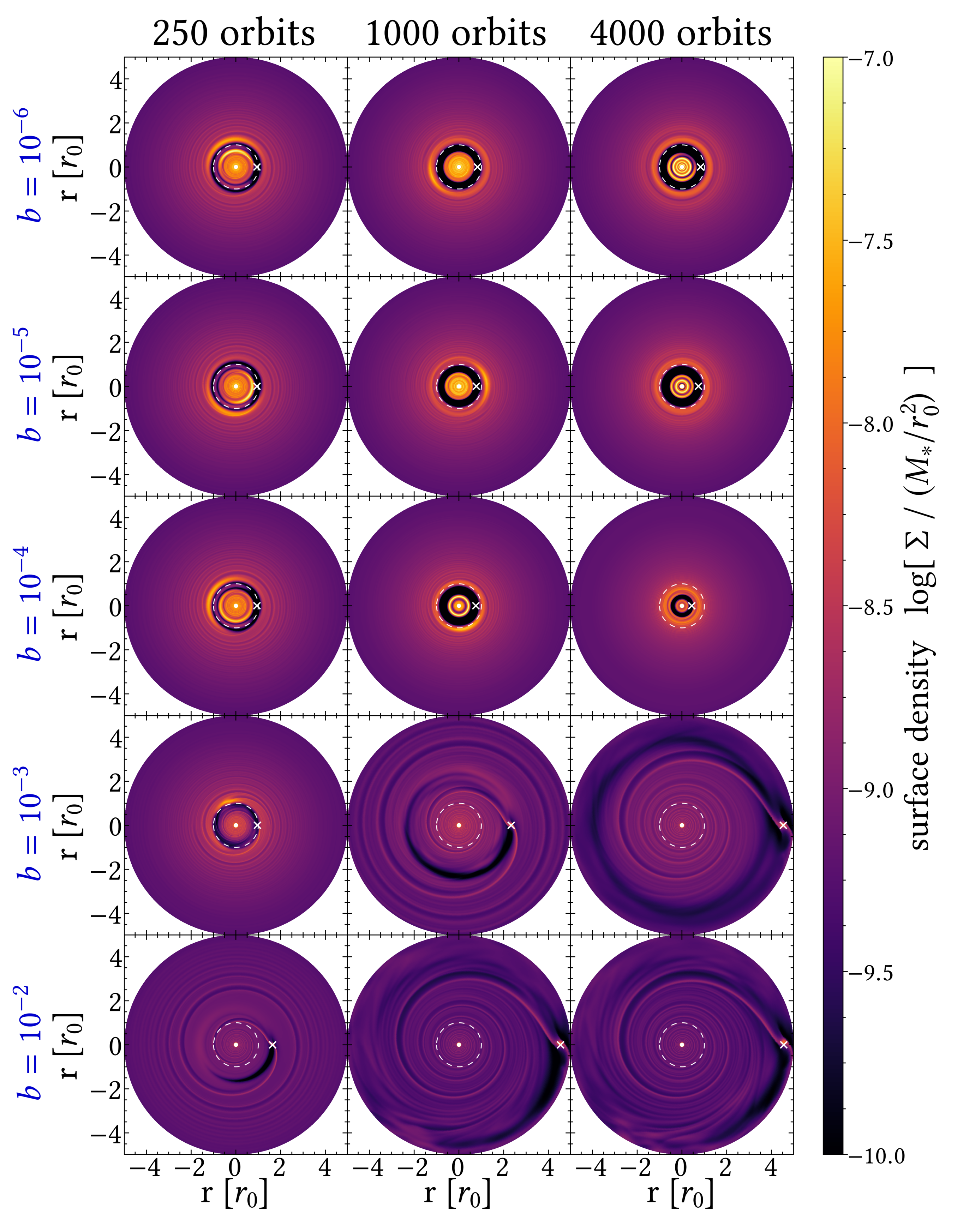}}
  \caption{\label{fig-windcase-msat-surfdens}Surface density maps for the
    wind-driven accretion case with a Saturn mass planet. The three columns are
    for different times. The five rows are for the different values of
    the mass loss parameter $b$. The dashed white circle marks the
    initial orbit of the planet. The cross denotes its current location.}
\end{figure*}

\begin{figure*}
  \centerline{\includegraphics[width=0.8\textwidth]{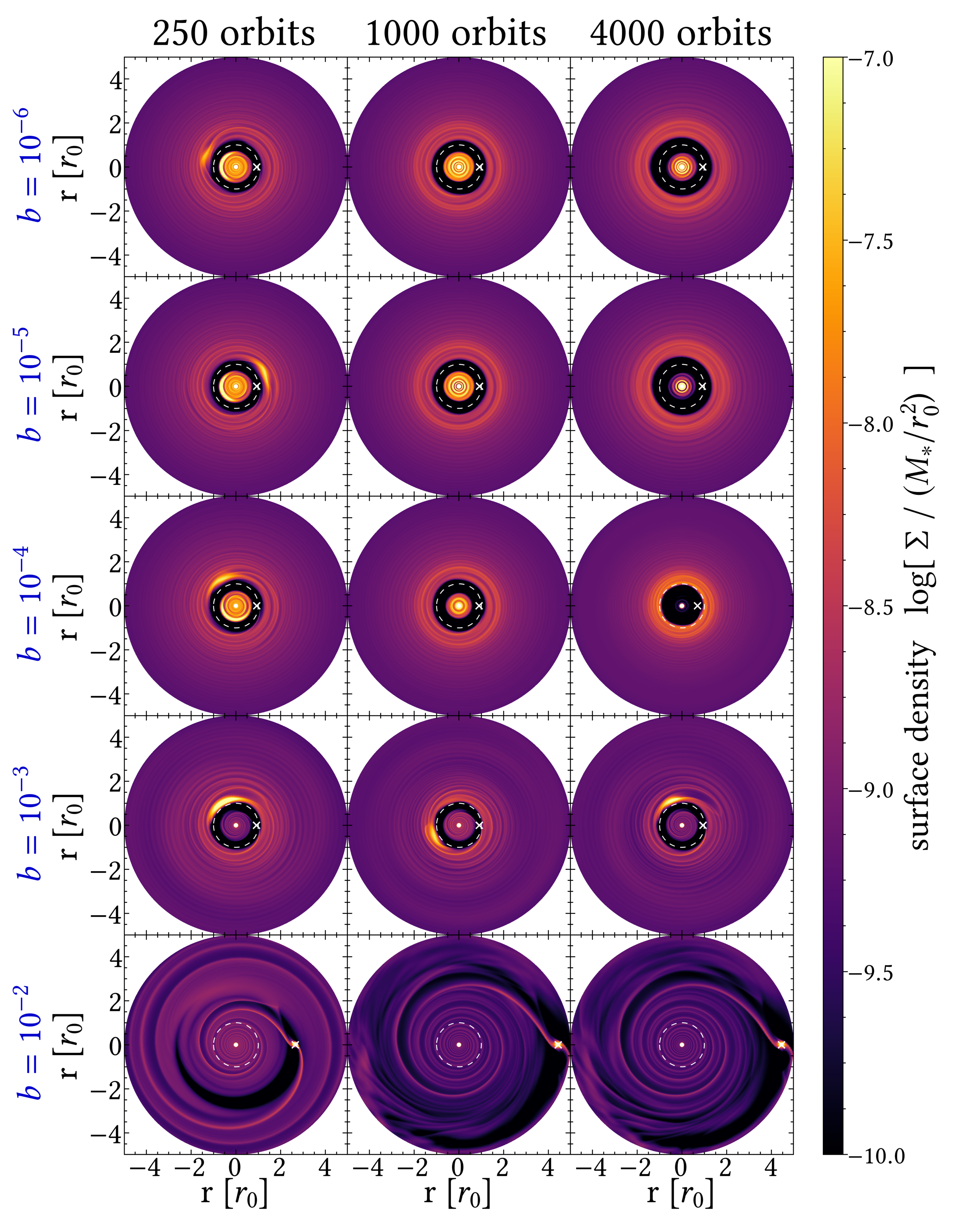}}
  \caption{\label{fig-windcase-mjup-surfdens}As Fig.~\ref{fig-windcase-msat-surfdens},
  but now for the Jupiter mass planet.}
\end{figure*}

Now we switch to the wind-driven accretion. We set $\lambda=2.25$ and vary
$b=10^{-6},\;10^{-5},\;10^{-4},\;10^{-3},\;10^{-2}$. The resulting disk surface
density structures are shown for the Saturn and Jupiter mass planet in
Fig.~\ref{fig-windcase-msat-surfdens} and
\ref{fig-windcase-mjup-surfdens}, respectively.

For low values of $b$ (the cases $b=10^{-6},\;10^{-5},\; 10^{-4}$) we see that
the planet opens a gap, as expected. In the outer edge of the gap, an elongated
vortex forms, which is a known effect for low-viscosity disks
\citep{2003ApJ...596L..91K, 2000ApJ...533.1023L, 2013A&A...553L...3A}.
One can also observe the
formation of secondary gaps in the inner disk, consistent with the findings of
\citet{2017ApJ...850..201B} and \citet{2017ApJ...843..127D} for low-viscosity
disks. These secondary rings take many orbits to form and are most pronounced
in the 4000 orbit panels.

For the Saturn mass planet models we find clear inward migration for
$b\le 10^{-4}$. The planet location (shown in the Figure with a cross) after 4000
orbits is well within the original orbit (shown in the Figure as the dashed
circle). For the Jupiter mass planet models the inward migration is weaker, but
for the case $b=10^{-4}$ it is nevertheless clearly seen
(Fig.~\ref{fig-windcase-mjup-surfdens}).

For large values of $b$ (the cases $b=10^{-3},\;10^{-2}$ for the Saturn mass
planet, and $b=10^{-2}$ for the Jupiter mass planet) we observe a totally
different behavior. The gap is opened only partially, and asymmetrically. While
the co-orbital region trailing the planet is evacuated, the co-orbital region
leading the planet still contains substantial amounts of gas. This leads to
a positive corotation torque, which adds angular momentum to the planetary
orbit. As a result, the planet rapidly migrates outward. We will discuss this
mechanism in more detail in Section~\ref{sec-discussion}.

\begin{figure*}
  \centerline{\includegraphics[width=0.95\textwidth]{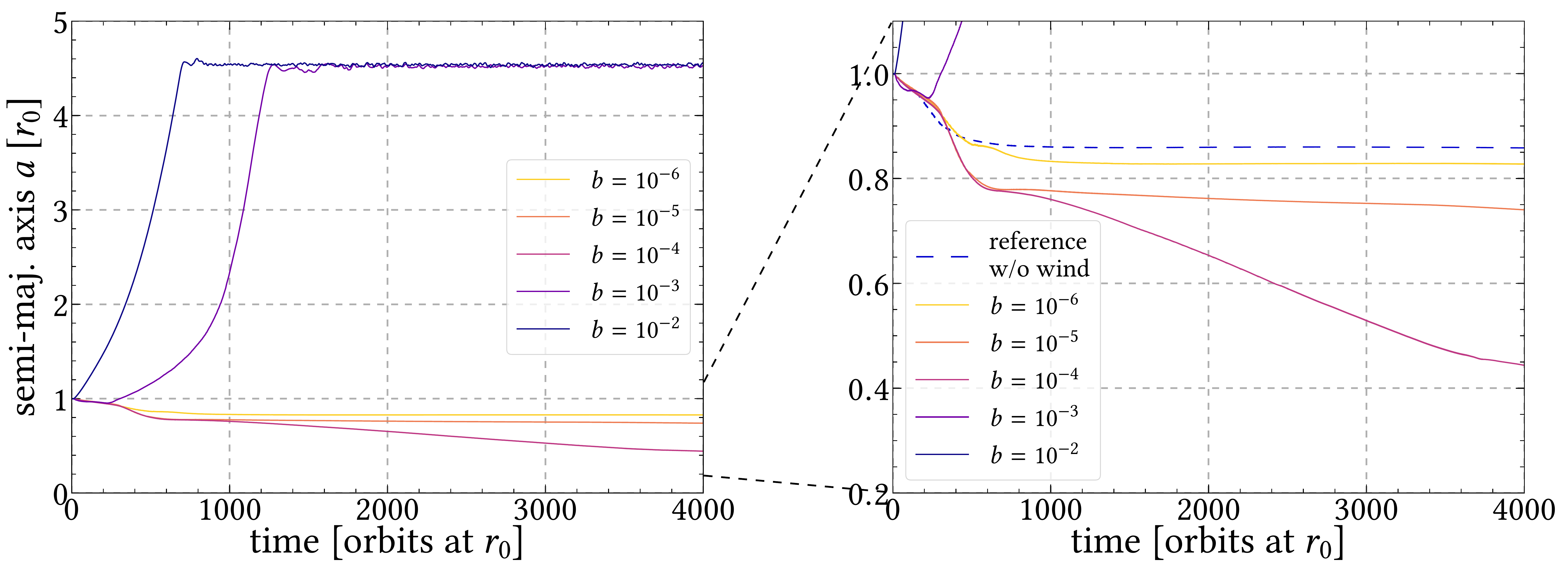}}
  \caption{\label{fig-windcase-sat-migration-b}The planetary migration for the
    wind-driven accretion case with Saturn mass planet, for the various values
  of $b$. Dashed line is the reference model without wind-driven accretion.}
\end{figure*}

\begin{figure*}
  \centerline{\includegraphics[width=0.95\textwidth]{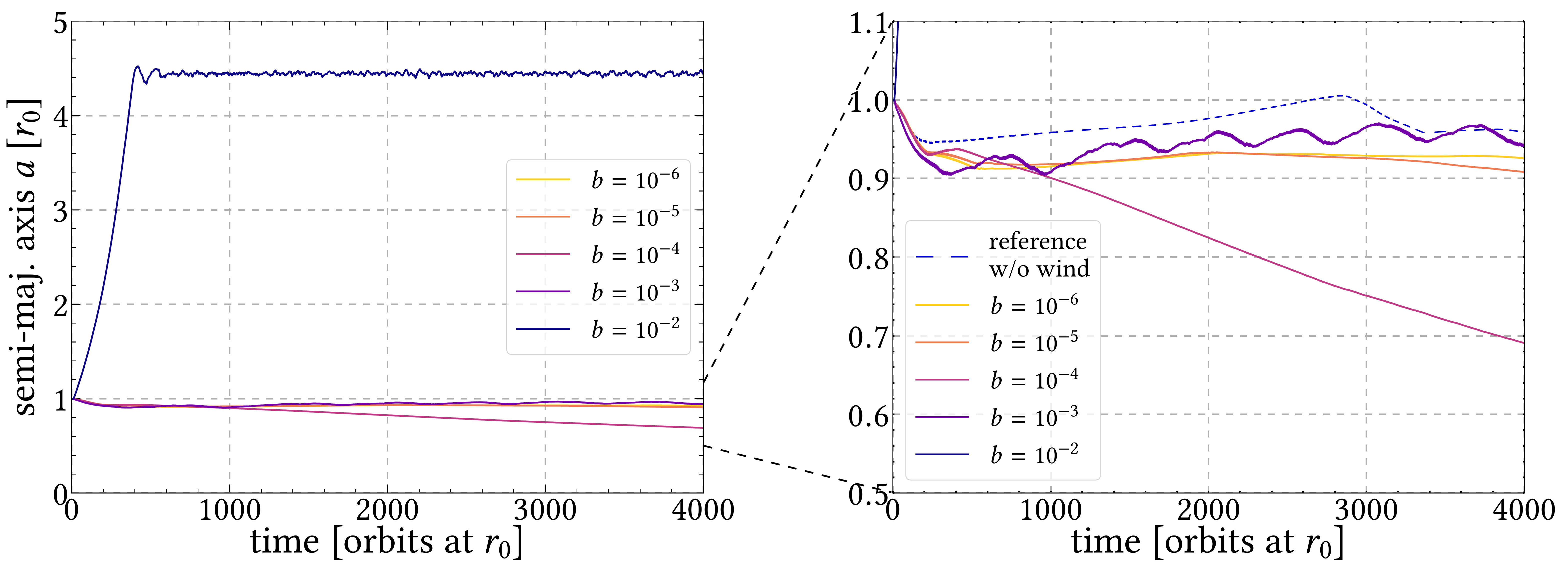}}
  \caption{\label{fig-windcase-jup-migration-b}As
    Fig.~\ref{fig-windcase-sat-migration-b}, but now for the Jupiter mass
    planet.}
\end{figure*}

In Figs.~\ref{fig-windcase-sat-migration-b} and
\ref{fig-windcase-jup-migration-b} we show the migration history of the planets
as a function of time. For some wind strengths the Saturn mass planet clearly
migrates inward or outward, in comparison to the reference case. It migrates
outward for $b=10^{-2}$ and $b=10^{-3}$, inward for $b=10^{-4}$, and slightly
inward for $b=10^{-5}$. In the case of $b=10^{-6}$, the planet initially
migrates further than in the reference case, but stalls after 1000 orbits. In
both outward migrating cases, the planet stops at $4.5\,r_0$, which is caused by
the outer boundary.

The Jupiter mass planet migrates outward only in the case of $b=10^{-2}$,
and inward for $b=10^{-4}$. In the case of $b=10^{-5}$ and $b=10^{-6}$, the
planet does not migrate significantly. Particularly, it does not migrate
outward as in the reference case without magnetic wind.

With a mass loss parameter of $b=10^{-3}$, the planet performs an unexpected
periodic inward and outward migration (see
Fig.~\ref{fig-windcase-jup-migration-b}). The surface density map of this model
(shown in Fig.~\ref{fig-windcase-mjup-surfdens} in the fourth row), shows
a stronger vortex in the outer disk than in the other simulations.
It does not vanish by time as in the other cases. We
have experimented with a higher resolution (1024$\times$1312 grid cells)
and found that while the vortex remains, the oscillation vanishes, and
that the planet migrates
somewhat further inward. It is therefore likely that the oscillations we observe
are a numerical artifact.

\begin{figure}
  \centerline{\includegraphics[width=0.5\textwidth]{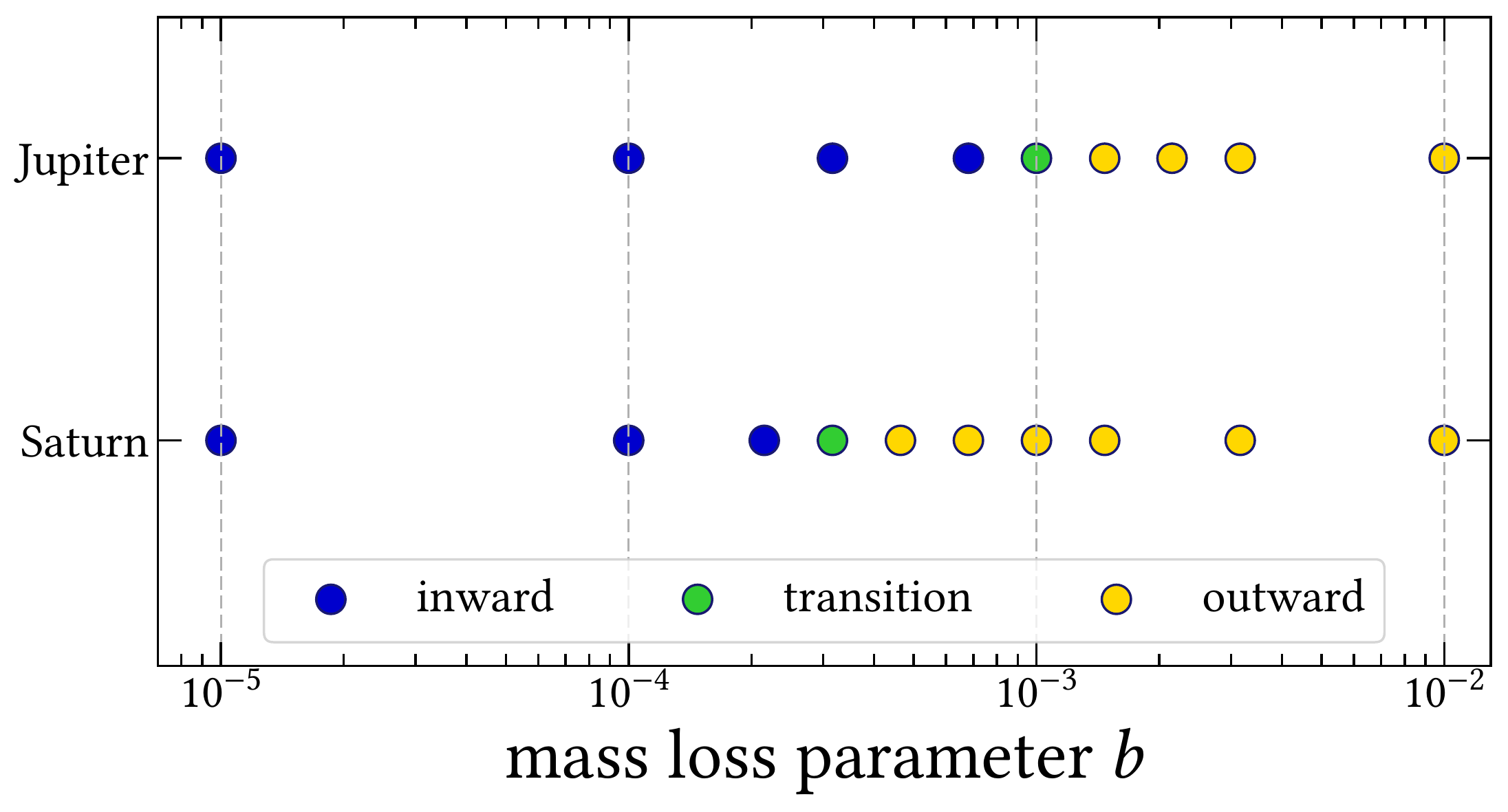}}
  \caption{\label{fig-threshold} Direction of migration in simulations
  with different mass loss parameters $b$ for both planet masses. The lever arm
  is $\lambda=2.25$ in all simulations.
  For the models marked as transition, no clear inward or outward migration was
  observed.}
\end{figure}

We suspect that this numerical artifact occurs when the simulation is
set up in the transition regime between inward and outward migration. In this
regime the planet is very sensitive to the different torques caused by the gas
surface density which could cause the oscillations.
To test this, we perform more simulations with the same setup as before,
except for the $b$-parameter.  We choose a finer logarithmic sampling for $b$ to cover
the transition regime for both planets, as shown in Fig. \ref{fig-threshold}.
We find for the Jupiter mass planet that the setup with $b=10^{-4}$ indeed turns
out to be the transition between inward and outward migration. For the Saturn planet,
the transition takes place at $b=3.16 \times 10^{-4}$ in which
we observe no migration of the Saturn planet and a light oscillation as in the
simulation with the Jupiter planet.

\begin{figure*}
  \centerline{\includegraphics[width=0.95\textwidth]{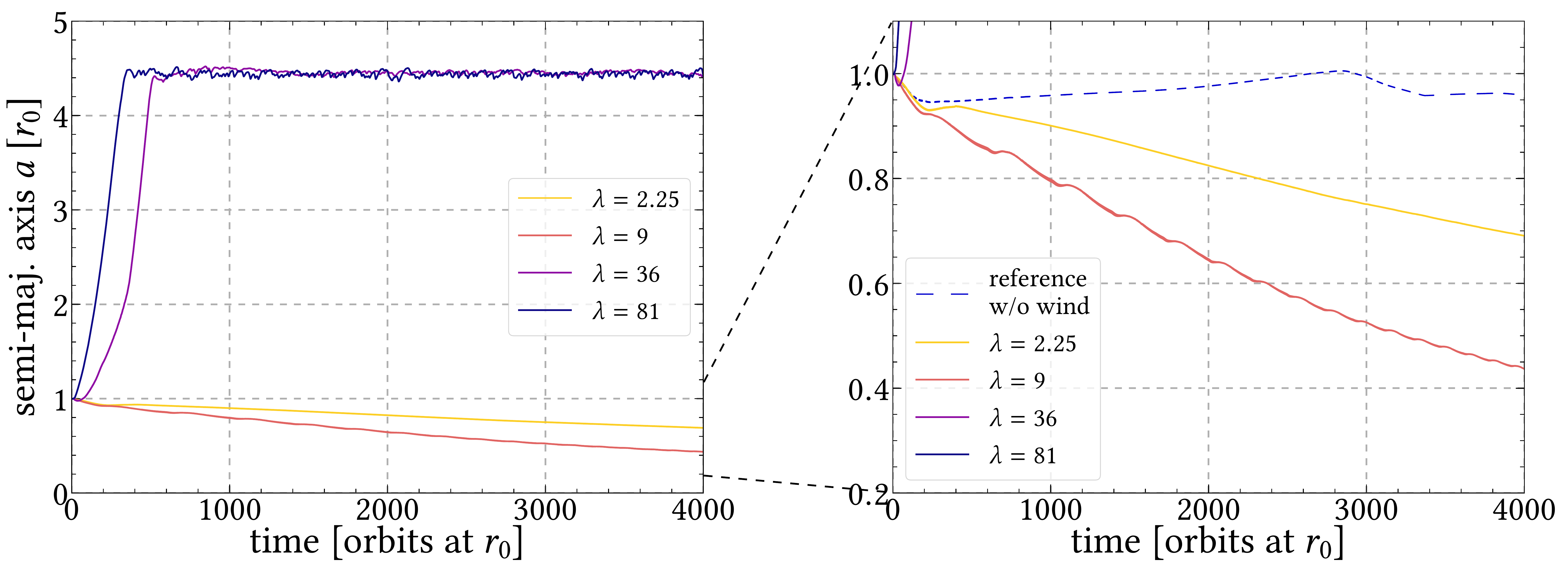}}
  \caption{\label{fig-windcase-sat-migration-leverarm}As
    Fig.~\ref{fig-windcase-jup-migration-b}, but now varying the
    lever arm $\lambda$, keeping $b=10^{-4}$ in all simulations. This
    is studied for the Jupiter mass
    planet.}
\end{figure*}

We also experiment with different values for the magnetic lever arm while we
keep the mass loss parameter at $b=10^{-4}$, as shown in
Fig.~\ref{fig-windcase-sat-migration-leverarm}. The
  torque on the disk is proportional to $b(\lambda-1)$, so that the effect
  of increasing $\lambda$ is expected to be similar to that of increasing
  $b$. However, increasing $\lambda$ does not increase the wind mass loss
  rate $\sigwind$, while increasing $b$ does. Hence, some differences
  in the results are expected. In these simulations,
we observe inward migration for low lever arms $\lambda=2.25,\;9$
and outward migration for high lever arms $\lambda=36,\;81$.
As in the investigation of the $b$-parameter, for low wind strength
the planet migrates inward while it migrates outward for high wind strengths.

\section{Discussion}\label{sec-discussion}

\subsection{Interpretation of the results}
Our models show that for low values of $b$ (weak wind loss) the planet migrates
inward, while for high values of $b$ (strong wind loss) the planet migrates outward.

In the low $b$ case the planet manages to open a gap and clear the co-orbital
region of its gas \citep[see, e.g.,][]{2012ARA&A..50..211K, 2018ApJ...861..140K}.
Therefore, there are no corotation torques acting on the planet and it
experiences only the Lindblad torques. The migration
is inward and to a certain extent similar to the standard type II
viscous migration, in the sense that gap formation plays a key
role in the migration process. We therefore call it type IIw migration. The torque by
the magnetic wind onto the disk gas causes the disk gas to gradually move
inward. Consequently, also the inner and outer edges of the gap move
inward. Suppose the planet stays on its position: then the outer edge moves
closer to the planet, while the inner edge moves further away from it. The
Lindblad torques from the outer gap edge then dominate over the Lindblad torques
from the inner gap edge, and thus push the planet inward to a new equilibrium
position.

\begin{figure}
  \centerline{\includegraphics[width=0.45\textwidth]{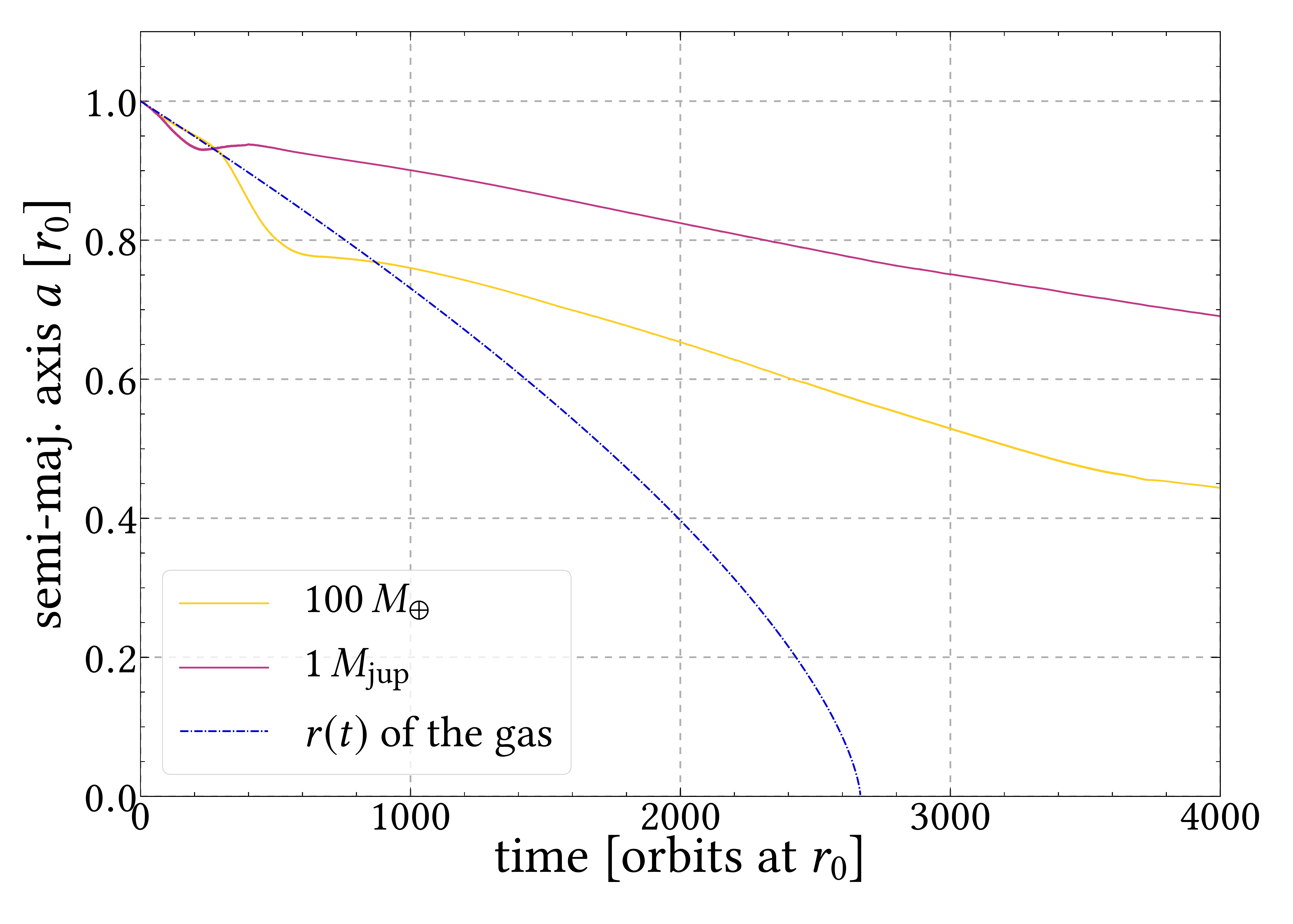}}
  \caption{\label{fig-compare-rmig-rgas}Migration of the planets compared
  to the gas motion, for the low $b$ case.}
\end{figure}

If the gas drift was the only process determining the migration, the migration
rate of the planet could be described by the radial velocity of the gas due to
the wind $v_{\mathrm{mig}}=v_r$. The planet would then be coupled to the disk,
and essentially follow the disk gas as it accretes inward. A closer inspection
of the results, however, shows that some gas leaks through the gap, making the
type IIw migration somewhat slower than the gas inward motion. This can be
seen in Fig.~\ref{fig-compare-rmig-rgas}. Similar effects are known
to occur in migration of gap-opening planets in viscous disks
\citep{2014ApJ...792L..10D, 2017A&A...598A..80D}.

It is interesting to note that, while type IIw migration seems similar to type
II migration, there is a fundamental difference between the two: In a viscous
disk, the gas motion can be both inward or outward. In particular the very outer
disk regions expand outward (the typical Lynden-Bell \& Pringle disk
evolution). This would lead to outward migration. In the case of wind-driven
accretion, the wind always removes angular momentum from the disk, and therefore
the gas always moves inward. Thus, type IIw also always has to point inward.

\begin{figure}
  \centerline{\includegraphics[width=0.45\textwidth]{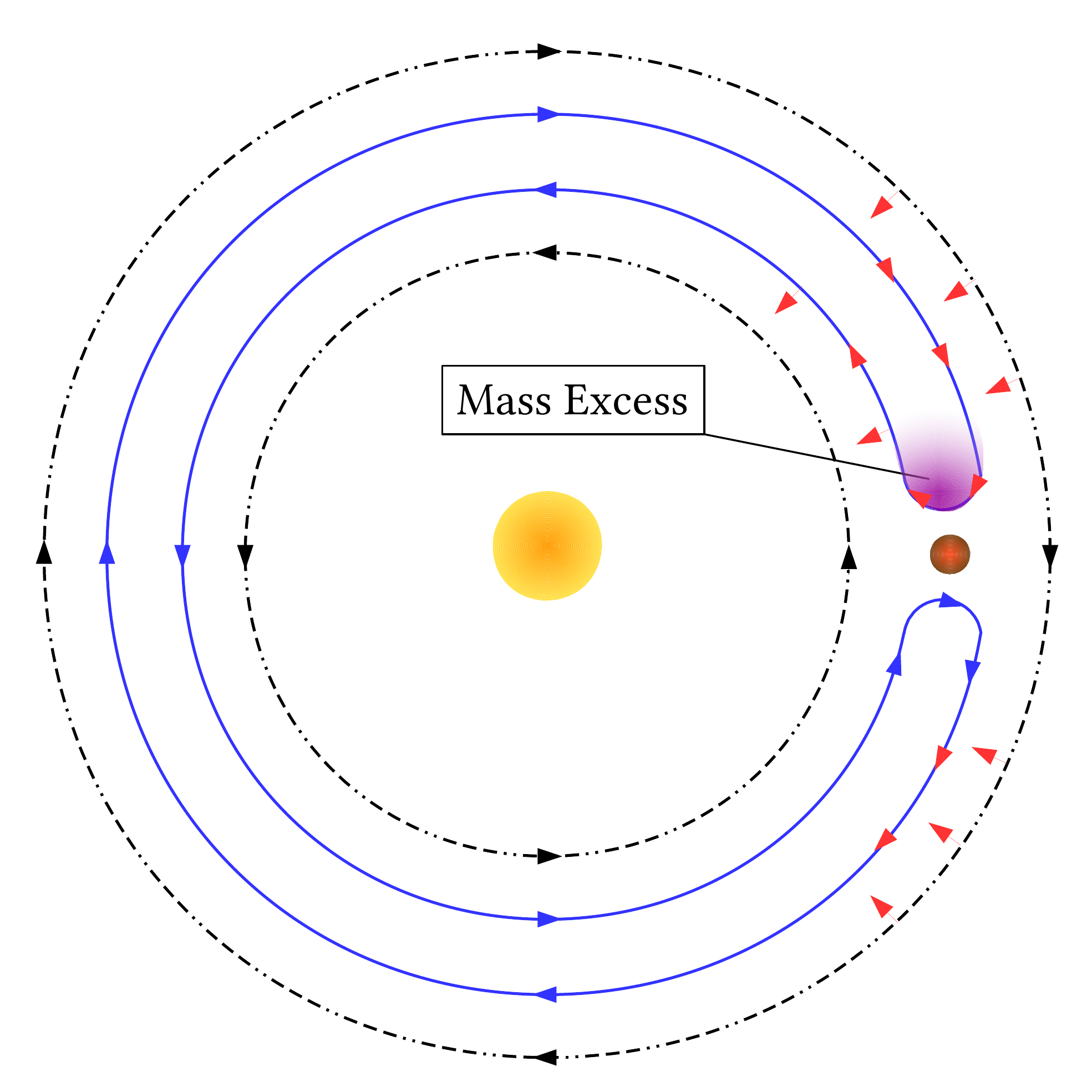}}
  \caption{\label{fig-cartoon}Cartoon of how the wind-driven inward motion of
    the gas through the co-orbital region leads to a strongly asymmetric mass
    distribution, with a mass excess in front of the planet and a mass deficit
    trailing the planet.}
\end{figure}

Now let us turn to the behavior for large $b$, where the planet migrates
outward at a high speed. In this case, the radial inward motion of the gas is
rapid enough that it can enter the horseshoe streamlines, as shown in
Fig.~\ref{fig-cartoon}. This leads to a strong asymmetry in the horseshoe
region. When gas enters the corotation region in front of the planet, it
performs a horseshoe orbit and moves toward the planet. At the U-turn point of
the horseshoe orbit an excess of mass occurs. While the gas turns its direction
at this point, it moves closer to the star, which means that it loses angular
momentum. This angular momentum is transferred to the planet, which in return
gains it. 

Behind the planet, a defect of mass occurs, because the horseshoe orbit
transports the gas away from the planet.
A key to maintaining this defect is that the gas that performed the U-turn in
front of the planet now finds itself close to the inner edge of the gap.  Due to
the rapid inward motion caused by the wind, this gas then quickly leaves the
co-orbital region again by entering the inner disk. This means that this gas
will not librate all the way to the back side of the planet, leaving this region
devoid of gas. In terminology of standard migration theory, one can say that
this process avoids saturation of the corotation torque.

Since only the U-turn in front of the planet is populated with gas, the planet
only gains angular momentum, not loses it. The planet thus migrates outward.
This outward migration mechanism is very similar to type III ``runaway
migration'' \citep{2003ApJ...588..494M}. In our case, however, the migration is
always outward, while in the case of type III migration, the direction of the
initial migration of the planet determines the run-away migration direction.
We call this type IIIw migration.

We estimate the parameters for the transition from type IIw migration
to type IIIw by comparing the libration timescale $\tau_\mathrm{lib}$, which is
the time a gas parcel takes to complete a horseshoe orbit, to the time it takes
to radially cross the horseshoe region driven by the magnetic wind, which we call
passing time $\tau_\mathrm{pass}$. The libration timescale is
\citep{2012ARA&A..50..211K}
\begin{equation}
\tau_\mathrm{lib} = \frac{8 \pi}{3 \Omega_\mathrm{K,p}} \frac{ r_\mathrm{p}}{\Delta r},
\end{equation}
where $r_\mathrm{p}$ is the distance from the planet to the star,
$\Omega_\mathrm{K,p}$ is the Kepler frequency at $r_\mathrm{p}$ and $\Delta r$ is
the half-width of the horseshoe region. We can estimate the passing time as
\begin{equation}
\tau_\mathrm{pass} = \frac{2\Delta r}{|{v_r(r_\mathrm{p})}|} = \frac{2\pi}{\Omega_\mathrm{K,p}\ b (\lambda -1)} \frac{\Delta r}{r_\mathrm{p}},
\end{equation}
using Eq. \ref{eq-radvelo-with-b}. Comparing the timescales results in
\begin{equation}
\frac{\tau_\mathrm{pass}}{\tau_\mathrm{lib}} = \frac{3}{4} \left(\frac{\Delta r}{r_\mathrm{p}}\right)^2 \frac{1}{b (\lambda -1)}.
\end{equation}
A shorter passing time and a longer libration time means that the gas is faster to
cross the corotation region than to complete the horseshoe orbit. This keeps the
corotation torque unsaturated and we expect outward migration.
The half-width of the horseshoe region is found to be
$\Delta~r~=~C(\epsilon)~r_\mathrm{p}~\sqrt{q/h}$, with $C(\epsilon)$ as factor of
order unity and $\epsilon$ as smoothing length, $q$ as the planet-to-star mass ratio
and $h$ as scale height \citep[see][]{2012ARA&A..50..211K}. We use this to define
a criterion $K = \tau_\mathrm{pass}/\tau_\mathrm{lib}$ for $C(\epsilon)=1$
\begin{equation}
K = \frac{3}{4} \frac{q}{h} \frac{1}{b (\lambda-1)}.
\end{equation}
In our simulations, we find that the transition regime between inward
and outward migration occurs for $K~\approx~10$. For $K~\gtrsim~10$ we find inward
migration and for $K~\lesssim~10$ outward migration.

\subsection{Caveats}
One caveat of our flat disk model is that we do not treat the possible vertical
stratification of the coupling of the magnetic fields to the gas. It is believed
that the well-shielded midplane regions of protoplanetary disks are ``dead'', in
the sense that magnetic fields decouple from the gas. The surface layers are
likely sufficiently ionized to be coupled to the field lines. This means that
the torque the magnetocentrifugal wind exerts onto the disk is only applied to
the very upper layers of the disk, not the full vertical extent of the
disk. We argue that this does not make a big difference for our model, because
by the conservation of angular momentum, the accretion rate driven by this
mechanism only depends on the torque. The only effect of the dead midplane zone
is that all of this wind-driven accretion will then have to be carried by the
surface layer material. For the present model it is irrelevant whether the
accretion is carried by the full column of the disk or only by the surface
layers.

Furthermore, our parametrization of the winds assume a constant lever
arm and mass loss parameter throughout the whole disk. A better approximation
would be a radial dependency of both parameters. The parameters are not well
known so far, therefore the used values are only estimates.

The resolution of our models is also an issue. Especially if the mass of the
planet is small, the width of the co-orbital region is narrow, requiring high
spatial resolution. We have already seen the effect of resolution on the results
in Section~\ref{sec-results}, when we discussed the case of $1\,M_{Jup}$ and
$b=10^{-3}$. 

The biggest caveat of our simplified approach is that our model does not set the
wind rate self-consistently. The wind rate and lever arm, and thereby the torque
onto the disk, have to be completely parametrized. Improvement requires a
detailed 3-D magnetohydrodynamic simulation of the driving of the wind.  Such
models exist \citep[see, e.g.,][]{2017A&A...600A..75B, 2017ApJ...835...59W,
2019ApJ...874...90W},
but they are costly. To calculate the effect on planetary migration, many
thousands of orbits have to be computed, which is a challenge for such
models. Nevertheless, in the future this is the path that has to be taken.

\subsection{Comparison to other work}
The effect of wind mass loss of the disk on type I migration has been studied
before by, e.g., \citet{2015A&A...584L...1O}. In that study, however, the main
effect was the change of the radial profile of the disk surface density
$\Sigma(r)$, which, using the standard type I migration rate formula
\citep[e.g.,][]{2011MNRAS.410..293P} leads to a modified migration rate.
In contrast, in our paper we do not study how the changed $\Sigma(r)$
profile affects the migration rate (though it is, in a way, obtained for
free), but instead focus on the computation of the torque itself.

In addition to wind-driven accretion as a replacement for the classic
$\alpha$-viscosity model, there may be other drivers of accretion. For instance,
\citet{2017MNRAS.472.1565M} showed that the presence of a vertical magnetic
field in the dead zone of a disk could, via the Hall effect, lead to the
formation of strong spiral-shaped magnetic fields in the plane of the
disk. These fields transport angular momentum within the disk and lead to
accretion, even though the disk is laminar. They call this situation a
magnetically torqued dead zone. When a planet is inserted in such a disk, this
can lead to similar asymmetric filling/depletion of the co-orbital region and
the corresponding type-III-like migration effects as we find in our paper
\citep{2017MNRAS.472.1565M, 2018MNRAS.477.4596M}.  Given that, in our model, we
need a vertical magnetic field to drive a wind, it is very well possible that
both effects act simultaneously.

\section{Conclusions}\label{sec-conclusion}
Our models show that the effect of magnetocentrifugal wind-driven accretion on
planet migration can be profound. For very strong winds, it can even lead to
rapid outward migration akin to type III migration (which we call type IIIw
migration). In this case, however, the direction of migration is deterministic:
it does not depend on an initial ``seed motion'' of the planet. The parameter
range for which we find rapid outward migration may occur is, however, coupled
to a rapid evolution of the disk. We find that lower mass planets are more prone
to the type IIIw effect, and thus more easily migrate outward. We speculate that
this type IIIw outward migration mechanism may be a possible origin of the
intermediate mass planets at large radii that are thought to be the cause of the
multi-ringed ALMA disks \citep{2018ApJ...869L..42H, 2018ApJ...869L..47Z}. These
planets would then have formed at much smaller semi-major axes, and then
efficiently transported outward to their final location.

\begin{acknowledgements}
 We thank Leonardo Krapp, Philipp Weber and Thomas Rometsch for assistance with
 the required modifications to FARGO3D, and Hubert Klahr for useful comments.
 We also thank the referee, John Chambers, for his constructive and helpful
 comments that substantially improved the paper.
 Part of this work was supported by DFG grant DU 414/18-1 and KL 650/27-1
 within the Priority
 Programme ``Exploring the Diversity of Extrasolar Planets'' (SPP 1992)
 and DFG research group FOR 2634 ``Planet Formation Witnesses and Probes:
 Transition Disks'' under grant DU 414/23-1 and KL 650/30-1. For the simulations
 performed on
 the BwForCluster BinAC, we also acknowledge support by the High Performance and
 Cloud Computing Group at the Zentrum f\"ur Datenverarbeitung of the University of
 T\"ubingen, the state of Baden-W\"urttemberg through bwHPC and the German Research
 Foundation (DFG) through grant no INST 37/935-1 FUGG.
\end{acknowledgements}

\begingroup
\bibliographystyle{aa}
\bibliography{ms}
\endgroup

\end{document}